\documentclass[runningheads]{llncs}

\usepackage[review,year=***,ID=***]{eccv}
\usepackage{eccvabbrv}

\usepackage{graphicx}
\usepackage{booktabs}
 \usepackage{multirow}
\usepackage{pifont}
\usepackage[accsupp]{axessibility}  % Improves PDF readability for those with disabilities.
\usepackage[draft]{changes}
\usepackage{amsmath}

\usepackage[pagebackref,breaklinks,colorlinks,citecolor=eccvblue]{hyperref}
\begin{document}

\title{Learning Correction Errors via Frequency-Self Attention for Blind Image Super-Resolution} 

\titlerunning{LCE}

\author {Haochen Sun \and
Yan Yuan \and
Lijuan Su \and
Haotian Shao}


\institute{Beihang University, China \\
\email{sulijuan@buaa.edu.cn, yuanyan@buaa.edu.cn}}

\maketitle

\begin{abstract}
  Previous approaches for blind image super-resolution (SR) have relied on degradation estimation to restore high-resolution (HR) images from their low-resolution (LR) counterparts. However, accurate degradation estimation poses significant challenges. The SR model's incompatibility with degradation estimation methods, particularly the Correction Filter, may significantly impair performance as a result of correction errors. In this paper, we introduce a novel blind SR approach that focuses on Learning Correction Errors (LCE). Our method employs a lightweight Corrector to obtain a corrected low-resolution (CLR) image. Subsequently, within an SR network, we jointly optimize SR performance by utilizing both the original LR image and the frequency learning of the CLR image. Additionally, we propose a new Frequency-Self Attention block (FSAB) that enhances the global information utilization ability of Transformer. This block integrates both self-attention and frequency spatial attention mechanisms. Extensive ablation and comparison experiments conducted across various settings demonstrate the superiority of our method in terms of visual quality and accuracy. Our approach effectively addresses the challenges associated with degradation estimation and correction errors, paving the way for more accurate blind image SR.
  \keywords{Blind image super-resolution \and learning correction errors \and frequency learning}
\end{abstract}

\section{Introduction}
\label{sec:intro}

Single image super-resolution (SISR) aims at restoring an HR image from an LR image. A classical degradation model can be expressed as:
\begin{equation}
\label{degradation}
{\bf{y}} = {({\bf{x}} * {\bf{k}})\downarrow_s} + {\bf{n}},
\end{equation}
where, $\mathbf{x}$ is the original HR image, $\mathbf{y}$ is the LR image after blurring, downsampling, and added noise, $*$ denotes convolution between kernel and HR image, $ \downarrow_s$ is bicubic downsample operation \cite{knab1979interpolation} by scale $s$, and $\mathbf{n}$ is the noise term such as additive white Gaussian noise. For blind SR, $\mathbf{k}$ is an unknown kernel. For classic SR, $\mathbf{k}$ is the unit impulse response.
\begin{figure}
    \centering
    \includegraphics[width=0.9\textwidth]{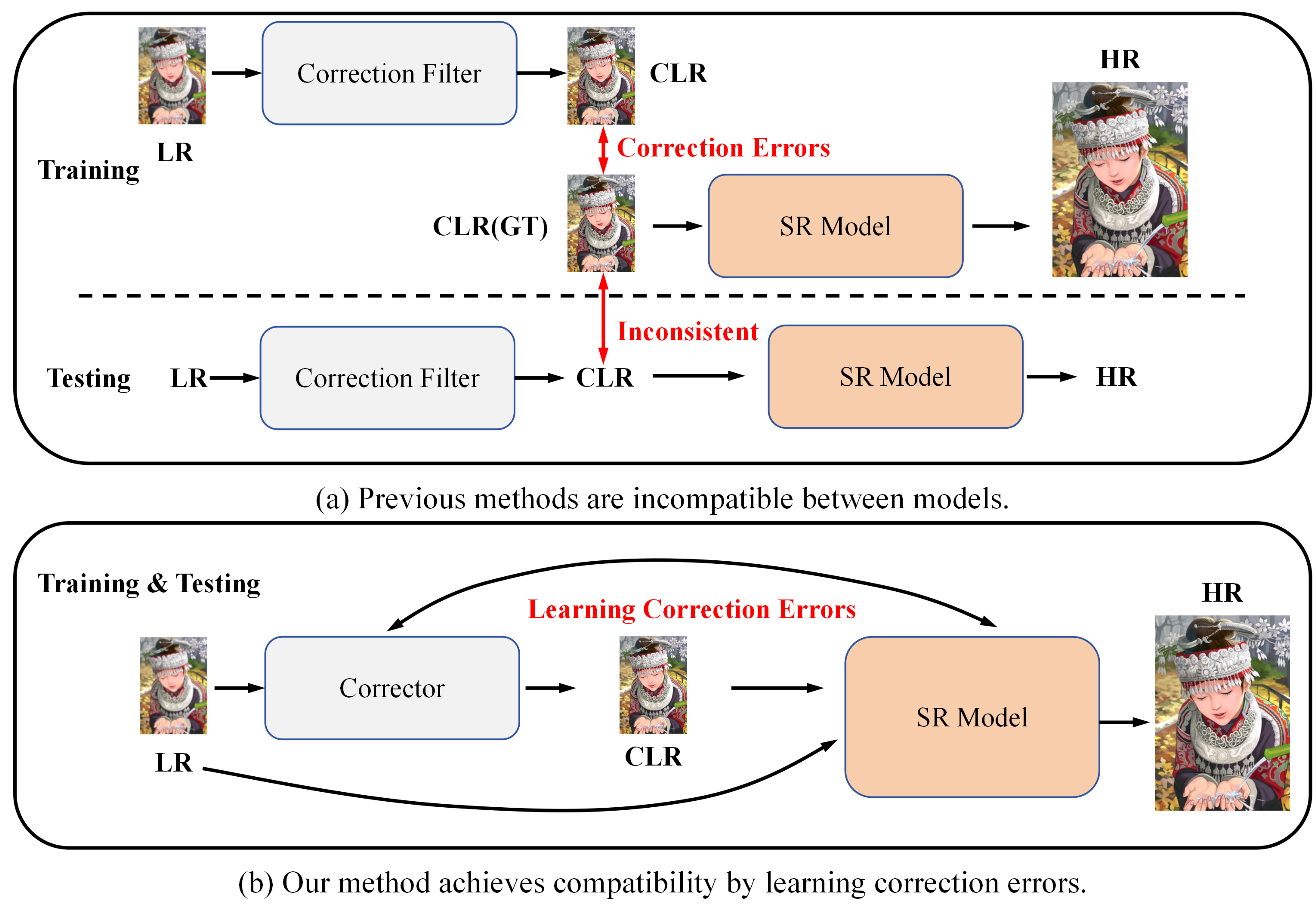}
    \caption{Difference between previous methods using Correction Filter and our method. (a) Previous methods are incompatible between models. (b) Our method achieves compatibility by learning correction errors. }
    \label{fig:first}
\vspace{-20pt}
\end{figure}

Previous methods \cite{KernelGAN,IKC,DAN,DCLS,DASR,hui2021learning,kim2021koalanet,AdaTarget,liang2021mutual, gong2023kernel, Correction_Filter} for blind SR problems often have two steps: \textbf{1):} degradation estimation. \textbf{2):} restoration using degradation estimation. Compared to classic SR methods \cite{EDSR, RCAN, HAN, SwinIR, HAT}, which only choose LR images as the network's input, blind SR approaches provide more degradation information for the SR network, expecting better restoration performance. Although many solutions in this pipeline have achieved outstanding results, some issues remain. First, the degradation estimation is not a simple task(i.e., blur kernel estimation), and additional estimation errors are introduced especially when the degradation is complex \cite{Blindsrsurvey}. Second, while another technical approach of a Correction Filter is accessible, it is not compatible with the SR model. Therefore, any correction errors will significantly undermine the overall restoration performance.

To take a step forward, we modify the blind SR pipeline by Learning Correction Errors to avoid the above issues. To be more precise, we first train a lightweight Corrector to estimate the CLR image, achieving high accuracy without any estimated kernel filter like Correction Filter\cite{Correction_Filter, DARSR}. The ground truth of the CLR image is the same as an LR image obtained by bicubic downsampling. To compensate for errors and artifacts of estimated CLR images, the original LR image is used as a reference along with the CLR image in an SR network to restore the final HR image. Since obvious frequency characteristics can be found in correction errors, frequency learning is used for the shallow feature extraction of CLR images. Moreover, any operation in the frequency domain is inherently global for the original feature space. Thus we introduce a Frequency-Self Attention Block (FSAB) to achieve better global representations than self-attention. It combines both window-based self-attention and frequency spatial attention.

Our work differs from previous methods of Correction Filter\cite{Correction_Filter, DARSR}. They all use a Correction Filter to get CLR images as the input for a classic SR model. However, a pretrained SR model ignores issues of correction errors when using estimated CLR images during inference and it will suffer a huge performance drop due to this incompatibility. On the other hand, our work focuses on learning correction errors to avoid this issue. The differences are shown in Fig. \ref{fig:first}. 

Furthermore, our work is also different from some low-level vision works which combine self-attention and frequency convolution\cite{swinfsr,zhu2023attention}. None of them tried spatial attention in the frequency domain, and they also did not explore any mixture of self-attention and frequency operation. In this paper, we present a two-branch architecture of both self-attention in feature space and spatial attention in the frequency domain to achieve better performance. We have conducted a detailed ablation study about these differences. 

\textbf{Contributions}: \textbf{(1)} We design a new blind SR architecture, which learns correction errors to improve the restoration ability of networks significantly. \textbf{(2)} We propose a novel Frequency-Self Attention Block that integrates self-attention and frequency spatial attention. \textbf{(3)} Extensive experiments on various degradation kernels demonstrate that our method leads to state-of-the-art performance in blind SR.

\section{Related Works}

\subsection{Deep Networks for Non-blind SR}
Since Dong \emph{et al.} \cite{SRCNN} initially introduced SRCNN based on deep convolution neural networks (CNN) to address the SISR task, a variety of approaches have been proposed to improve the performance by modifying architectures and changing training approaches \cite{VDSR, ESPCN, DBPN, SRresnet, Dense,mei2021, esrgan, zssr, metazssr}. For example, residual structure \cite{resnet} or dense residual structure \cite{densenet} have been proposed to make it feasible to train very deep networks. To discover the dependencies among different channels, the attention mechanism was introduced in \cite{RCAN, HAN}. Recently Transformer \cite{transformer} has achieved great success in many high-level vision tasks \cite{cvt, VIT,chu2021twins,swin, uniformer}, many Transformer-based methods \cite{VRT, EDT,cao2021video, Uformer, SwinIR, HAT} have been proposed for image SR task to get competitive results. However, if the actual degradation model deviates from the pre-assumed model (i.e., bicubic downsampling), the restoration quality of all these methods will suffer significantly.

\subsection{Deep Networks for Blind SR}
For the blind SR task, degradation kernels are unknown. A typical approach includes two steps. They first perform kernel estimation, and then use the estimated kernel and the LR image to complete SR. To utilize the recurrence property of nature images, KernelGAN \cite{KernelGAN} employed an Internal-GAN network trained on a set of specific images. FKP \cite{FKP} proposed a flow-based kernel prior to generating blur kernels. DASR \cite{DASR} employed a contrastive learning strategy to learn the implicit representation of various kernels. Gu \emph{et al.} \cite{IKC} iteratively corrected kernel estimation and image restoration using three split networks. DAN \cite{DAN} adopted an end-to-end network to optimize the estimator and restorer alternately. Husseins \emph{et al.} \cite{Correction_Filter} used a Correction Filter to obtain the corrected LR image matching the one obtained by bicubic downsampling, and DARSR \cite{DARSR} further introduced a spatial variant Correction Filter through self-supervision. Luo \emph{et al.} \cite{DCLS} reformulated the degradation model and applied a deep constrained least square (DCLS) filter to get the clean features. KDSR \cite{KDSR} employed knowledge distillation to implicitly estimate degradation.

\subsection{Frequency Learning}
Frequency analysis has been commonly used in image processing for years. But for deep learning, in 2020, Chi \emph{et al.} \cite{FFC} first introduced Fast Fourier Convolution.  In 2021, Rao \emph{et al.} \cite{globalfilter} proposed a learnable global filter in the frequency domain. Guibas \emph{et al.} \cite{AFNO} proposed a Fourier token mixer for Transformer to improve performance. In 2023, Lin \emph{et al.} \cite{frequencyfilter} proposed a deep frequency filter for domain generalization.  

Some works \cite{swinfsr,zhu2023attention} of low-level vision also employ a combination of self-attention and frequency convolution. Chu \emph{et al.} \cite{chu2023rethinking} proposed Unbiased Fast Fourier Convolution to achieve better performance and robust training. However, they did not explore the combination of self-attention and spatial attention in the frequency domain.

\section{Methods}
\label{sec:blind}

\subsection{Motivation}
\label{3.1}
For simplicity, following the method in \cite{DCLS}, we reformulate Eq. (\ref{degradation})  as: 
\begin{equation}
\label{reformulated}
    \mathbf{y} = \mathbf{x}\downarrow_s*\mathbf{k}_{l} + \mathbf{n}
\end{equation}
% and
\begin{equation}
    \mathbf{k}_{l} = \mathcal{F}^{-1}\left(\frac{\mathcal{F}\left(\left(\mathbf{x}*\mathbf{k}\right)\right\downarrow_s)}{\mathcal{F}\left(\mathbf{x}\downarrow_s\right)}\right),
\end{equation}
where \begin{math}\mathcal{F}\end{math} represents the Discrete Fourier transform and \begin{math} \mathcal{F}^{-1}\end{math} is its inverse. \(\mathbf{x}\downarrow_s\) is the groundtruth of CLR image downsampled by a bicubic kernel (anti-aliasing) \cite{antialiasing}. 
$\mathbf{k}_{l}$ is exactly the downsampled version of the original $\mathbf{k}$ regardless of spatial aliasing.

For blind SR, the goal is to find the HR image $\mathbf{x}$ from the LR image $\mathbf{y}$. Many previous methods \cite{IKC,DAN,DCLS} use a two-step method of kernel estimation to solve this problem. They first estimated $\mathbf{k}$ or $\mathbf{k_l}$, then the kernel and $\mathbf{y}$ are fed into an SR model to restore the HR image. However, If we want to estimate 
\begin{figure}[]
    \centering
    \includegraphics[width=0.9\textwidth]{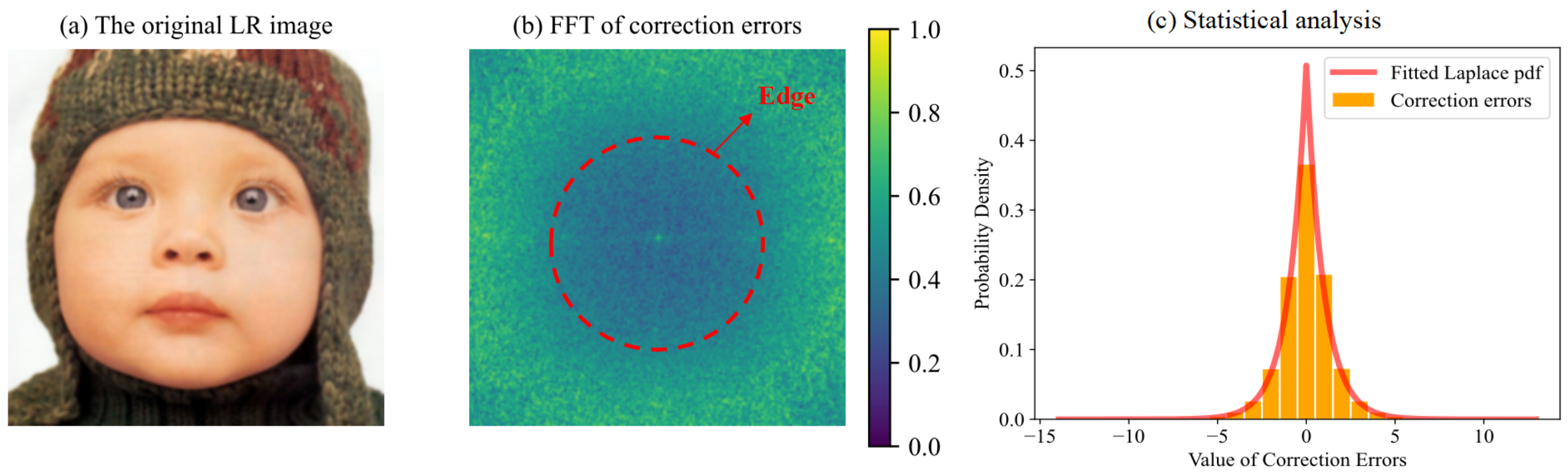}
    \caption{Analysis of correction errors.}
    \label{fig:frequency analysis}
    \vspace{-20pt}
\end{figure}
kernel, neural networks should involve the prior of \(\mathbf{x}\downarrow_s\). This problem is highly ill-posed. Kernel estimation is tough especially when the degradation becomes complicated\cite{Blindsrsurvey}.

Another approach desires to utilize existing SR models designed for the classic SR task. They use a Correction Filter to correct \(\mathbf{y}\) to \(\mathbf{x}\downarrow_s\). Then, the estimated CLR image is taken as the input of an SR model trained for the classic SR task:
\begin{equation}
\label{equation 6}
\begin{cases}
     \mathbf{x}\downarrow_{s,e} = CF\left(\mathbf{y}\right), \\ \mathbf{x} = SR_{pre}\left(\mathbf{x}\downarrow_{s,e}\right),
\end{cases}
\end{equation}
where $\mathbf{x}\downarrow_{s,e}$ denotes estimated CLR images, $CF$ is a Correction Filter, and $SR_{pre}$ is an SR model trained under classic bicubic downsampling.
There are two main issues involved.  First, the estimated CLR image inevitably has correction errors that have not been learned by the classic SR model. The disparity of this input distribution degrades the performance of SR restoration. The strict definition of correction errors can be formulated simply as:
\begin{equation}
    \mathbf{L_c} =  \mathbf{x}\downarrow_s - \mathbf{x}\downarrow_{s,e}, 
\end{equation}
where, $\mathbf{L_c}$ denotes correction errors in CLR images. Second, the complete information of the original LR image cannot be exploited for the SR model due to the replacement of the CLR image, which is a significant loss. Our goal is to effectively tackle these two issues.

\subsection{Formulation}
\label{3.2}
 We analyze correction errors in CLR images using the Corrector of section \ref{3.3}. The visual result is shown in Fig. \ref{fig:frequency analysis}. We use L1 Loss for training, which inherently assumes that the errors follow a Laplace distribution. The actual statistical analysis in Fig. \ref{fig:frequency analysis} confirms this. Correction errors in CLR images are distributed in the high-frequency range. It can be modeled as an image with additive high-frequency Laplace noise. According to the analysis above, we reformulated $\mathbf{x}\downarrow_{s,e}$ as:
 \begin{equation}
     \mathbf{x}\downarrow_{s,e} = \mathbf{x}\downarrow_s + \mathbf{n_c},
 \end{equation}
where $\mathbf{n_c}$ is the additive high-frequency noise. The previous methods of Correction Filter\cite{Correction_Filter, DARSR} use $\mathbf{x}\downarrow_s$ for the training of an SR model and utilize $\mathbf{x}\downarrow_{s,e}$  for testing, which impairs the performance of restoration. To tackle two problems in section \ref{3.1}, we reformulated the pipeline as:
\begin{equation}
\label{equation 6}
\begin{cases}
     \mathbf{x}\downarrow_{s,e} = C\left(\mathbf{y}\right), \\ \mathbf{x} = \mathrm{arg}  \:\underset{\mathbf{x}}{\mathrm{min}}   \|\mathbf{y} -(\mathbf{x}*{\mathbf{k}})\downarrow_s\| + \|\mathbf{x}\downarrow_{s,e} -\mathbf{x}\downarrow_s\| + \phi\left(\mathbf{x}\right),
\end{cases}
\end{equation}
where $C$ is a Corrector to get $\mathbf{x}\downarrow_{s,e}$, and $\phi\left(\mathbf{x}\right)$ is the prior term for $\mathbf{x}$. In the second formula, we perform joint optimization targeting both $\mathbf{y}$ and $\mathbf{x}\downarrow_{s,e}$. The first term utilizes $\mathbf{y}$ to avoid unnecessary loss of information and the second learns correction errors in $\mathbf{x}\downarrow_{s,e}$ in the way of dealing with high-frequency noise.

\subsection{Architecture}
\label{3.3}
\begin{figure*}[t]
\centering
\includegraphics[width=1\textwidth]{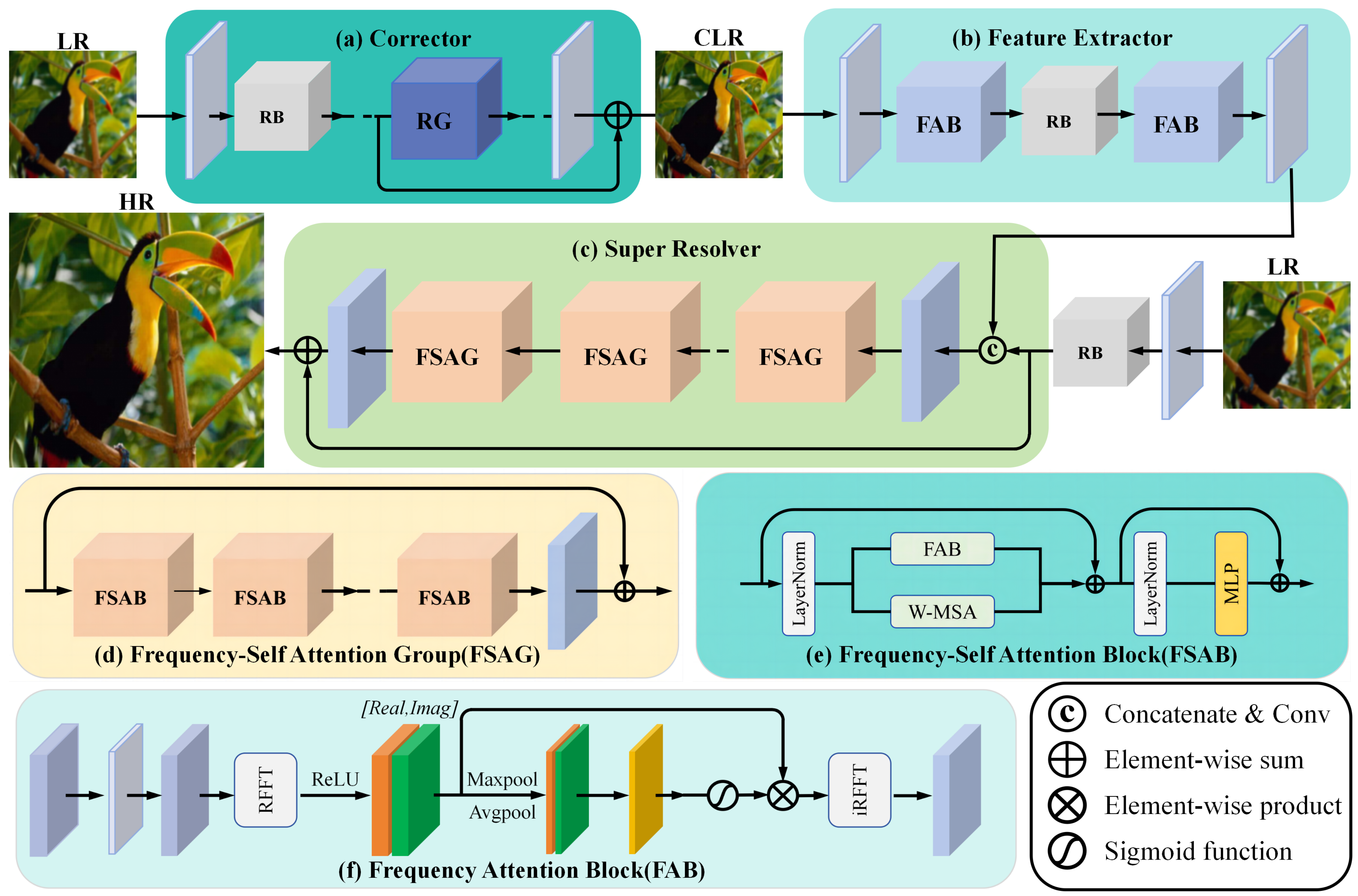}
\caption{The overall architecture of the method LCE, consisting of a lightweight Corrector, a Feature Extractor for CLR images, and a Super Resolver.}
\vspace{-10pt}
\label{Fig_archict}
\end{figure*}

As depicted in Fig. \ref{Fig_archict}, we design a simple but effective architecture according to Eq. (\ref{equation 6}). The whole structure mainly consists of a Corrector, a Feature Extractor, and a Super Resolver. A lightweight Corrector transforms the original LR image into a CLR image. Next, the CLR image with correction errors is fed into a Feature Extractor to extract its unique shallow features with frequency learning. The features of the original LR image are extracted by a simple Resblock (RB). The two features are then concatenated together to generate a fused feature map for joint optimization. Then, the Super Resolver extracts deep features and upsamples them to create the final SR image. The long skip connection is used for both the Corrector and the Super Resolver.
\begin{figure}
    \centering
    \includegraphics[width=0.9\textwidth]{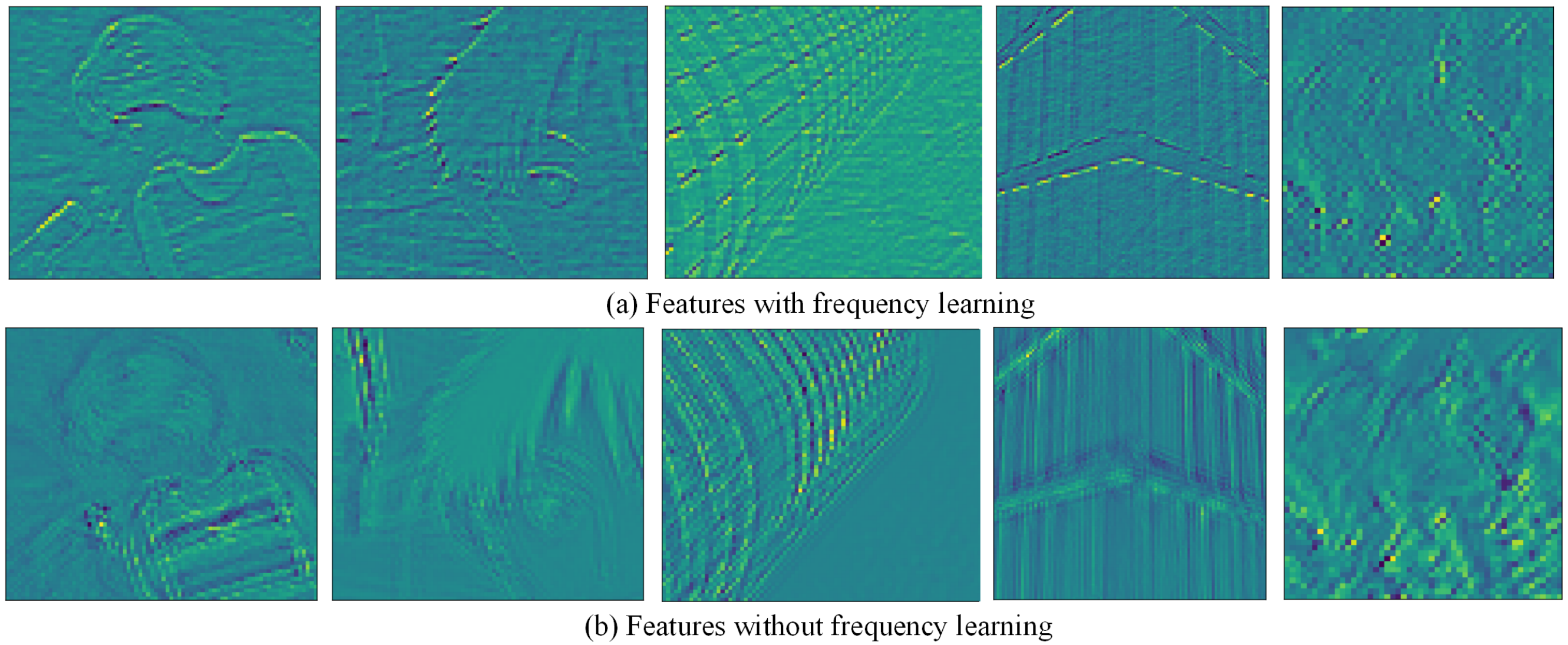}
    \caption{Comparison of features extracted by different methods.}
    \label{fig:feature}
    \vspace{-10pt}
\end{figure}

\subsubsection{Corrector and CLR Feature Extractor.}
\label{3.2.1}
The previous Correction Filter\cite{Correction_Filter, DARSR} still needs the estimated kernels as the kernel filter. Here, our Corrector gives up kernel estimation and uses a simple and lightweight architecture. Specifically for a given input LR image \(I^{LR} \in\mathbb{R}^{H\times{}W\times{}C_{in}}\), a single \(7\times{}7\) convolution layer and 2 RBs are used to transform it to shallow features \(f_0 \in \mathbb{R}^{H\times{}W\times{}C}\). Then residual groups (RG), composed of residual channel attention blocks (RCAB) \cite{RCAN}, are used to extract deep features \(f_i \in \mathbb{R}^{H\times{}W\times{}C}\) from \(f_{i-1}\):
\begin{equation}
    f_i = \mathrm{RG_i}\left(f_{i-1}\right), i = 1, 2,\dotsc, N.
\end{equation}

Upsampling is not required for correction, so after a total of \(N\) RGs, two convolution layers are used to create the final CLR image:
\begin{equation}
    I^{CLR} = F_{conv2}\left(F_{conv1}\left(f_N\right)+f_0\right).
\end{equation}

We use a Feature Extractor for the estimated CLR image to get its shallow features. As shown in Fig. \ref{fig:frequency analysis}(b), correction errors have obvious high-frequency characteristics. It indicates that applying modules of frequency learning to shallow feature extraction tends to be useful for noise filtering. In particular, two Frequency Attention Blocks (FAB) and an RB are employed. We further conduct a comparison of features extracted by different methods in Fig . \ref{fig:feature}. Features extracted with frequency learning are clearer and sharper, demonstrating its effectiveness.

\subsubsection{Super Resolver.}
\label{3.2.2}
As discussed in section \ref{3.2}, 
for joint optimization of CLR images with correction errors and the original LR images, we do not use a pretrained SR model for the classic SR task. Instead, we take features of both CLR images and original LR images as the input. Specifically, For \(I^{CLR}, I^{LR} \in \mathbb{R}^{H\times{}W\times{}C_{in}}\), their shallow features \(F_0^{CLR}, F_0^{LR} \in \mathbb{R}^{H\times{}W\times{}C}\) are extracted respectively. Only a simple RB is used for feature extraction of LR images where no correction errors are involved. 

Then two features are concatenated together and compressed to create a mixed feature map \(F_0 \in \mathbb{R}^{H\times{}W\times{}C}\):
\begin{equation}
    F_0 = H_{fuse}\left(cat\left(F_0^{CLR}, F_0^{LR}\right)\right).
\end{equation}

We extract deep features via Frequency-Self Attention Group (FSAG) :
\begin{equation}
    F_{i} = H_{FSAG}\left(F_{i-1}\right), i = 1, 2,\dotsc, N.
\end{equation}

As shown in Fig. \ref{Fig_archict}(d), a Frequency-Self Attention Group (FSAG) contains several FSABs, a $3\times{3}$ convolution layer, and a skip connection. 
The strategy of shift window-based self-attention in Swin Transformer\cite{swin} reduces computational complexity and brings better efficiency compared with ViT\cite{VIT}, but it also limits the utilization of global information\cite{HAT}. On the other hand, any operation in the frequency domain is inherently global for the original feature space. Motivated by the Frequency Filter\cite{frequencyfilter} for domain generalization and previous attention design \cite{RCAN, cbam, HAN}, as illustrated in Fig. \ref{Fig_archict}(e), we design a two-branch mixture in FSAB to acquire a strong mapping ability of self-attention and global interaction of frequency attention simultaneously. For a input feature  \(X\), we can get the out feature \(X_{out}\) using FSAB:
\begin{align}
\begin{split}
    X_1 &= \mathrm{LN}\left(X\right), \\
    X_2 &= \mathrm{W\text{-}MSA}\left(X_1\right) + \alpha \mathrm{FAB}\left(X_1\right) + X_1, \\
    X_{out} &= \mathrm{MLP}\left(LN\left(X_2\right)\right) + X_2,
\end{split}
\end{align}
where, \(\rm{FAB}\) denotes a Frequency Attention Block, and \(\rm{W\text{-}MSA}\) denotes shift window-based self-attention \cite{SwinIR} performed in the original feature space. \(\rm{MLP}\) is a multi-layer perception. LN denotes the LayerNorm layer. For calculating \(\rm{W\text{-}MSA}\), the strategy of shifted window partitioning \cite{swin} is used among a series of FSABs, and we embed each pixel as a token. \(\alpha\) is a coefficient to adjust the weights of output between \(\rm{W\text{-}MSA}\) and \(\rm{FAB}\). Considering FABs at different positions have unique relations with \(\rm{W\text{-}MSA}\), we set \(\alpha\) as a learnable parameter instead of a fixed hyperparameter.

\subsubsection{Frequency Attention Block (FAB).}
\label{3.2.3}
 Here we introduce our instantiation of FAB in which efficiency is fully considered. As shown in  Fig. \ref{Fig_archict}(f), for an input feature $Y\in \mathbb{R}^{H\times{}W\times{}C}$, we first perform two consecutive convolutions to get $Y_1$. However, embedding dimensions in Transformer are much larger than the number of channels in common CNN architecture, so we employ a channel compression between two convolution operations\cite{HAT} to improve efficiency. Then we perform Real Fast Fourier Transform (RFFT) and ReLU function on $Y_1$ to get $Y_F\in \mathbb{R}^{H\times{}\left(\lfloor W /2 \rfloor + 1\right)\times{}2C}$. The transform above can be formulated as:
\begin{equation}
Y_F=\mathrm{ReLU}\left(\mathrm{RFFT}\left(F_{conv2}\left(F_{conv1}\left(Y\right)\right)\right)\right).
\end{equation}
The features after RFFT retain half spatial size due to conjugate symmetry. This property also reduces computational complexity. Then, max-pooling for the real part and average-pooling for the imaginary part along the dimension of the channel are conducted to get $Y_{max}\in \mathbb{R}^{H\times{}\left(\lfloor W /2 \rfloor + 1\right)\times{}1}$ and  $Y_{avg}\in \mathbb{R}^{H\times{}\left(\lfloor W /2 \rfloor + 1\right)\times{}1}$:
\begin{equation}
     Y_{2} = \mathrm{Maxpool}\left(Y_F.r\right), Y_{3} = \mathrm{Avgpool}\left(Y_F.i\right).
\end{equation}
\begin{table}[]
\centering
\caption{Ablation study results on our proposed Learning Correction Errors (LCE).}
\label{LCE}
\setlength{\tabcolsep}{5pt}
\begin{tabular}{cclclcccccc}
\hline
\multirow{2}{*}{Methods} & \multicolumn{2}{c}{\multirow{2}{*}{Corrector}} & \multicolumn{2}{c}{\multirow{2}{*}{LCE}} & \multicolumn{2}{c}{BDS100 \cite{BSD100}} & \multicolumn{2}{c}{Urban100 \cite{Urban100}} & \multicolumn{2}{c}{Manga109 \cite{Manga109}} \\
 & \multicolumn{2}{c}{} & \multicolumn{2}{c}{} & PSNR & SSIM & PSNR & SSIM & PSNR & SSIM \\ \hline
case1 & \multicolumn{2}{c}{\ding{55}} & \multicolumn{2}{c}{\ding{55}} & 32.10 & 0.8913 & 31.74 & 0.9201 & 38.53 & 0.9738 \\
case2 & \multicolumn{2}{c}{\ding{51}} & \multicolumn{2}{c}{\ding{55}} & 32.05 & 0.8914 & 31.66 & 0.9185 & 38.05 & 0.9736 \\
case3 & \multicolumn{2}{c}{\ding{51}} & \multicolumn{2}{c}{\ding{51}} & \textbf{32.14} & \textbf{0.8916} & \textbf{31.99} & \textbf{0.9233} & \textbf{38.81} & \textbf{0.9749} \\
\hline
\end{tabular}
\end{table}
\begin{table}[ht]
\centering
\caption{Ablation study on our proposed FAB.}
\label{table:fsab}
\setlength{\tabcolsep}{4pt}
\begin{tabular}{ccccccccc}
\hline
\multirow{2}{*}{Structure} & \multicolumn{2}{c}{baseline} & \multicolumn{2}{c}{w/CAB} & \multicolumn{2}{c}{w/FFC} & \multicolumn{2}{c}{w/FAB} \\
 & PSNR & SSIM & PSNR & SSIM & PSNR & SSIM & PSNR & SSIM \\ \hline
BSD100 \cite{BSD100} & 27.68 & 0.7323 & 27.66 & 0.7309 & 27.67 & 0.7321 & \textbf{27.69} & \textbf{0.7326} \\
Urban100 \cite{Urban100} & 26.57 & 0.7951 & 26.55 & 0.7935 & 26.58 & 0.7952 & \textbf{26.70} & \textbf{0.7984} \\
Manga109 \cite{Manga109} & 31.01 & 0.9132 & 30.99 & 0.9124 & 31.00 & 0.9135 & \textbf{31.14} & \textbf{0.9149} \\ \hline
\#Params	&\multicolumn{2}{c}{-}&\multicolumn{2}{c}{17.6K}&\multicolumn{2}{c}{186.9K}&		\multicolumn{2}{c}{\textbf{15.8K}} \\ \hline	
\end{tabular}
\vspace{-10pt}
\end{table}
After a great reduction of channels,  a $7\times7$ convolution layer as well as a sigmoid function is used to get a spatial attention map $Y_{map}\in \mathbb{R}^{H\times{}\left(\lfloor W/2 \rfloor + 1\right)\times{}1}$. The attention map is multiplied element-wise with the original feature $Y_F$. After that, inverse Real Fast Fourier Transform (iRFFT) is performed to get the feature $Y_4$ in the original space. It can be formulated as:
\begin{align}
    Y_4 &= \mathrm{iRFFT}\left(Y_{F} \odot F_{conv3}\left([Y_2, Y_3]\right)\right).
\end{align}

In the whole process, we avoid performing convolution operations when the number of channels is large. This is one of the main differences compared to the design in CNN architecture \cite{frequencyfilter}.

\section{Experiments and Discussion}

\subsection{Experiment Setup and Datasets}

Following methods in \cite{DAN, DCLS}, we use a total of 3450 frames of 2K images from DIV2K \cite{DIV2K} and Flickr2K \cite{Flicker2K} as the ground truth of HR images in the training datasets. We conduct experiments on both isotropic Gaussian kernels and anisotropic Gaussian kernels. For isotropic Gaussian kernels, we follow the setting of \cite{IKC}. Five well-known benchmarks: Set5 \cite{Set5}, Set14 \cite{Set14}, BSD100 \cite{BSD100}, Urban100 \cite{Urban100}, Manga109 \cite{Manga109} are used to generate testsets. For anisotropic Gaussian kernels, the setting in \cite{KernelGAN} is followed, we use the DIV2KRK testset used in \cite{KernelGAN} for evaluation. To obtain additional evaluation results for the anisotropic kernel setting, we also create a new testset named Urban100RK by using Urban100. Using the same setting as DIV2KRK, we randomly generate 100 anisotropic Gaussian kernels to blur and downsample HR images in Urban100.

The Corrector employs 4 RGs, each composed of 10 RCABs, and the channel number is set to 64. For the Super Resolver, we set the window size to 16. The channel number is set to 144. The FSAG, FASB, and attention head numbers
\begin{table}[!t]
\setlength{\tabcolsep}{5pt}
\centering
\caption{Comparison with the commonly used method of kernel estimation.}
\label{kernel estimation}
\begin{tabular}{ccccccc}
\hline
\multirow{2}{*}{Methods} & \multicolumn{2}{c}{BSD100 \cite{BSD100}} & \multicolumn{2}{c}{Urban100 \cite{Urban100}} & \multicolumn{2}{c}{Manga109 \cite{Manga109}} \\
 & PSNR & SSIM & PSNR & SSIM & PSNR & SSIM \\ \hline
LCE & \textbf{32.14} & \textbf{0.8916} & \textbf{31.99} & \textbf{0.9233} & \textbf{38.81} & \textbf{0.9749} \\
kernel estimation & 32.06 & 0.8908 & 31.21 & 0.9137 & 38.29 & 0.9734 \\ \hline
\end{tabular}
\end{table}
\begin{table}[!t]
\setlength{\tabcolsep}{10pt}
\centering
\caption{Ablation study on RG/RB in Corrector.}
\begin{tabular}{cccc|c} \hline
RG/RB & 2/5 & 4/10 & 4/20 & Correction Filter \cite{Correction_Filter} \\ \hline
PSNR & 49.88 & 50.30 & 50.31 & 32.58 \\ \hline
\end{tabular}
\label{tab 3}
\vspace{-10pt}
\end{table}
are set to 6. We use L1 loss for both of them. We first train the Corrector. After the Corrector is fixed, we then train the Super Resolver for stability. The initial weighting coefficient $\alpha$ of FAB is set as 0.01. PSNR and SSIM \cite{SSIM} on the Y channel are used for quantitative evaluation. More training details can be found in the supplementary.

\subsection{Ablation Study}
\subsubsection{Effectiveness of Learning Correction Errors.}
\label{4.2.1}
We conduct experiments to demonstrate the effectiveness of Learning Correction Errors. We train the Super Resolver for three scenarios: (1) Using only pairs of HR and LR images as the baseline of case 1. (2) Using Corrector but not Learning Correction Errors (scheme in \cite{Correction_Filter, DARSR}) as case 2. (3) Using HR, LR, and Learning Correction Errors in CLR images (ours) as case 3. For ×2 SR under isotropic kernels, the quantitative results are reported on all three large datasets in Tab. \ref{LCE}. Performance for all datasets can be further enhanced by our proposed LCE. Using the Corrector without Learning Correction Errors produces worse performance than the baseline. It indicates the inconsistency between training and testing severely impaired the performance of restoration.

\subsubsection{Effectiveness of Frequency Attention Block.}
We conduct the ablation experiment to prove the effectiveness of our proposed FAB. Considering channel attention block (CAB) \cite{RCAN} is a common block used in SR, we train networks for four scenarios: (1) Using only the W-MSA as the baseline. This scenery is similar to SwinIR\cite{SwinIR}. (2) Using W-MSA and standard CAB to comprise a two-branch architecture. (3) Using W-MSA and Fast Fourier Convolution (FFC). FFC directly employs frequency convolution instead of frequency spatial attention. (4) Using W-MSA and our proposed FAB to comprise a two-branch mixture architecture (FSAB). The quantitative results for scale 4 on isotropic kernel are reported in Tab. \ref{table:fsab}. We find that our proposed FAB outperforms CAB and FFC in terms of restoration performance with fewer parameters.
\begin{table}[]
\caption{Quantitative comparison with other methods on datasets with isotropic kernel. The top two results are marked in \color{red}{red}, \color[HTML]{0070C0}blue\color{black}.}
\label{tab 4}
\centering
\fontsize{7pt}{\baselineskip}\selectfont
\resizebox{1\textwidth}{!}{ 
\begin{tabular}{cccccccccccc}
\hline
 &  & \multicolumn{2}{c}{Set5 \cite{Set5}} & \multicolumn{2}{c}{Set14 \cite{Set14}} & \multicolumn{2}{c}{BSD100 \cite{BSD100}} & \multicolumn{2}{c}{Urban100 \cite{Urban100}} & \multicolumn{2}{c}{Manga109 \cite{Manga109}} \\
\multirow{-2}{*}{Method} & \multirow{-2}{*}{Scale} & PSNR & SSIM & PSNR & SSIM & PSNR & SSIM & PSNR & SSIM & PSNR & SSIM \\ \hline
Bicubic &  & 28.82 & 0.8577 & 26.02 & 0.7634 & 25.92 & 0.7310 & 23.14 & 0.7258 & 25.60 & 0.8498 \\
DASR\emph{\textcolor[RGB]{255,165,0}{[CVPR 2021]}}\cite{DASR} & & 37.02 & 0.9502 & 32.61 & 0.8957 & 31.59 & 0.8813 & 30.26 & 0.9015 & 36.20 &  0.9686 \\
DARSR \emph{\textcolor[RGB]{255,165,0}{[ICCV 2023]}}\cite{DARSR} &  & {26.42} & {0.7982} & {26.86} & 0.7816 & {26.48} & {0.7483} & {25.23} & {0.7975} & {27.27} & {0.8808} \\
IKC\emph{\textcolor[RGB]{255,165,0}{[CVPR 2019]}}\cite{IKC} &  & 37.19 & 0.9526 & 32.94 & 0.9024 & 31.51 & 0.8790 & 29.85 & 0.8928 & 36.93 & 0.9667 \\
DANv1\emph{\textcolor[RGB]{255,165,0}{[NeurIPS 2020]}}\cite{DAN} &  & 37.34 & 0.9526 & 33.08 & 0.9041 & 31.76 & 0.8858 & 30.60 & 0.9060 & 37.23 & 0.9710 \\
DANv2\emph{\textcolor[RGB]{255,165,0}{[arXiv 2021]}}\cite{DANv2} &  & 37.60 & 0.9544 & 33.44 & 0.9094 & 32.00 & 0.8904 & 31.43 & 0.9174 & 38.07 & 0.9734 \\
DCLS\emph{\textcolor[RGB]{255,165,0}{[CVPR 2022]}}\cite{DCLS} &  & \color[HTML]{0070C0}{37.63} & {\color[HTML]{FF0000} {0.9554}} & 33.46 & {\color[HTML]{0070C0} 0.9103} & \color[HTML]{0070C0}{32.04} & 0.8907 & \color[HTML]{0070C0}{31.69} & \color[HTML]{0070C0}{0.9202} & \color[HTML]{0070C0}{38.31} & {\color[HTML]{0070C0} 0.9740} \\
KDSR-L\emph{\textcolor[RGB]{255,165,0}{[ICLR 2023]}}\cite{KDSR} &  & {37.47} & {0.9549} & \color[HTML]{0070C0}{33.48} & {0.9098} & 32.00 &\color[HTML]{0070C0}{0.8908} & 31.50 & 0.9186 & {37.38} & 0.9727 \\
LCE(ours) & \multirow{-9}{*}{×2} & {\color[HTML]{FF0000} {37.79}} & {\color[HTML]{0070C0} {0.9553}} & {\color[HTML]{FF0000} {33.70}} & {\color[HTML]{FF0000} {0.9112}} & {\color[HTML]{FF0000} {32.14}} & {\color[HTML]{FF0000}{0.8916} } & {\color[HTML]{FF0000} {31.99}} & {\color[HTML]{FF0000} {0.9233}} & {\color[HTML]{FF0000} {38.81}} & {\color[HTML]{FF0000} {0.9749}} \\ \hline
Bicubic &  & 24.57 & 0.7108 & 22.79 & 0.6032 & 23.29 & 0.5786 & 20.35 & 0.5532 & 21.50 & 0.6933 \\
DASR\emph{\textcolor[RGB]{255,165,0}{[CVPR 2021]}}\cite{DASR} & & 31.60 & 0.8824 &28.15 & 0.7595 & 27.36 & 0.7186 & 25.35 & 0.7538 & 29.80 & 0.8929 \\
IKC\emph{\textcolor[RGB]{255,165,0}{[CVPR 2019]}}\cite{IKC} &  & 31.67 & 0.8829 & 28.31 & 0.7643 & 27.37 & 0.7192 & 25.33 & 0.7504 & 28.91 & 0.8782 \\
DANv1\emph{\textcolor[RGB]{255,165,0}{[NeurIPS 2020]}}\cite{DAN} &  & 31.89 & 0.8864 & 28.42 & 0.7687 & 27.51 & 0.7248 & 25.86 & 0.7721 & 30.50 & 0.9037 \\
DANv2\emph{\textcolor[RGB]{255,165,0}{[arXiv 2021]}}\cite{DANv2} &  & 32.00 & 0.8885 & 28.50 & 0.7715 & 27.56 & 0.7277 & 25.94 & 0.7748 & 30.45 & 0.9037 \\
DCLS\emph{\textcolor[RGB]{255,165,0}{[CVPR 2022]}}\cite{DCLS} &  & {\color[HTML]{0070C0} 32.12} & 0.8890 & 28.54 & {0.7728} & {27.60} & {0.7285} & {\color[HTML]{0070C0} 26.15} & {0.7809} & {30.86} & {\color[HTML]{0070C0} 0.9086} \\
KDSR-L\emph{\textcolor[RGB]{255,165,0}{[ICLR 2023]}}\cite{KDSR} &  & 32.11 & \color[HTML]{FF0000}{0.8933} & {\color[HTML]{0070C0} 28.68} & \color[HTML]{FF0000}{0.7867} & \color[HTML]{0070C0}{27.64} & \color[HTML]{0070C0}{0.7300} & \color[HTML]{0070C0}{26.15} & {\color[HTML]{0070C0}0.7830} & \color[HTML]{0070C0}{30.99} & 0.9069 \\
LCE(ours) & \multirow{-7}{*}{×4} & {\color[HTML]{FF0000} {32.35}} & {\color[HTML]{0070C0} {0.8930}} & {\color[HTML]{FF0000} {28.69}} & {\color[HTML]{0070C0} {0.7761}} & {\color[HTML]{FF0000} {27.69}} & {\color[HTML]{FF0000} {0.7326}} & {\color[HTML]{FF0000} {26.70}} & \multicolumn{1}{l}{{\color[HTML]{FF0000} {0.7984}}} & {\color[HTML]{FF0000} {31.14}} & {\color[HTML]{FF0000} {0.9149}} \\ \hline 
\end{tabular}}
\vspace{-20pt}
\end{table}

\subsubsection{Comparison with Kernel Estimation.}
Kernel estimation is a common degradation estimation method in Blind SR\cite{zssr, IKC, KernelGAN, DANv2, DCLS}. We conduct the ablation study to explore its effect compared with Learning Correction Errors.  
Here, the network DDLK \cite{DCLS}, which is specifically designed for kernel estimation, is utilized. When we have the estimated kernel, we extract its shallow features and concatenate them with features from LR to generate a mixed feature, which is then fed into the Super Resolver, the same as that of LCE. The method above is representative since it is similar to the kernel stretching strategy commonly used in Blind SR \cite{IKC, DANv2, SRMD}. The quantitative results are provided in Tab. \ref{kernel estimation} for ×2 SR with isotropic kernels. With the same main body as Super Resolver, our LCE outperforms the commonly used method of kernel estimation. It demonstrates the superiority of our proposed method of LCE.

\subsubsection{Effects of the Size of Corrector.}
We conduct an ablation study to explore the effect of model size on correction results. The variables are the numbers of RG/RB, and the result is reported on Set5 for ×2 SR using the isotropic kernel. We find a tiny network (\emph{i.e.}, RG/RB: 2/5) can achieve a PSNR of 49.88 dB for this task. Further increasing the model size can improve performance but the effect will be marginal. However, as shown in Tab. \ref{tab 3}, the Correction Filter has a worse PSNR than our Corrector. The reason is that the HR is degraded by the composite of a bicubic kernel and a Gaussian kernel. It leads to problems and ill-conditional issues when using methods like Fourier transform in Correction Filter to decompose convolution.
\begin{table}[htbp]
\renewcommand{\arraystretch}{0.9}
\caption{Quantitative comparison with other methods with anisotropic kernel on DIV2KRK \cite{KernelGAN} and Urban100RK that we created. The top two results are marked in \color{red}{red}, \color[HTML]{0070C0}blue\color{black}.}
\label{tab 5}
\centering
\setlength{\tabcolsep}{7pt}
\begin{tabular}{cccccc}
\hline
 &  & \multicolumn{2}{c}{DIV2KRK \cite{KernelGAN}} & \multicolumn{2}{c}{Urban100RK} \\
\multirow{-2}{*}{Method} & \multirow{-2}{*}{Scale} & PSNR & SSIM & PSNR & SSIM \\ \hline
Bicubic &  & 28.73 & 0.8040 & 22.14 & 0.6053 \\
IKC\emph{\textcolor[RGB]{255,165,0}{[CVPR 2019]}}\cite{IKC} &  & 31.44 & 0.8793 & 23.07 & 0.6570 \\
DANv1\emph{\textcolor[RGB]{255,165,0}{[NeurIPS 2020]}}\cite{DAN} &  & 32.56 & 0.8997 & 27.75 & 0.8440 \\
DANv2\emph{\textcolor[RGB]{255,165,0}{[arXiv 2021]}}\cite{DANv2} &  & 32.58 & 0.9048 & 28.14 & 0.8533 \\
DCLS\emph{\textcolor[RGB]{255,165,0}{[CVPR 2022]}}\cite{DCLS} &  & \color[HTML]{0070C0}32.75 & \color[HTML]{0070C0}0.9094 & \color[HTML]{0070C0} 28.39 & \color[HTML]{0070C0}0.8594 \\
KDSR\emph{\textcolor[RGB]{255,165,0}{[ICLR 2023]}}\cite{KDSR} &  & {32.73} & {0.9050} & {23.21} & {0.6502} \\ 
LCE(ours) & \multirow{-7}{*}{×2} & {\color[HTML]{FF0000} {32.98}} & {\color[HTML]{FF0000} {0.9107}} & {\color[HTML]{FF0000} {29.37}} & {\color[HTML]{FF0000} {0.8770}} \\ \hline
Bicubic &  & 25.33 & 0.6795 & 21.36 & 0.5596 \\
DASR\emph{\textcolor[RGB]{255,165,0}{[CVPR 2021]}} \cite{DASR} & & 28.10 & 0.7687 & 23.96 & 0.6939 \\
IKC\emph{\textcolor[RGB]{255,165,0}{[CVPR 2019]}} \cite{IKC}&  & 27.70 & 0.7668 & 23.58 & 0.6802 \\
DANv1\emph{\textcolor[RGB]{255,165,0}{[NeurIPS 2020]}}\cite{DAN}&  & 27.55 & 0.7582 & 23.44 & 0.6718 \\
DANv2\emph{\textcolor[RGB]{255,165,0}{[arXiv 2021]}}\cite{DANv2}&  & 28.74 & 0.7893 & 24.93 & 0.7384 \\
DCLS\emph{\textcolor[RGB]{255,165,0}{[CVPR 2022]}}\cite{DCLS}&  & \color[HTML]{0070C0}28.99 & \color[HTML]{0070C0}0.7946 & \color[HTML]{0070C0}25.53 & \color[HTML]{0070C0}0.7594 \\
KDSR\emph{\textcolor[RGB]{255,165,0}{[ICLR 2023]}}\cite{KDSR} & & {28.23} & {0.7757} & {24.66} & {0.7253} \\
LCE(ours) & \multirow{-7}{*}{×4} & {\color[HTML]{FF0000} {29.20}} & {\color[HTML]{FF0000} {0.8022}} & {\color[HTML]{FF0000} {26.02}} & {\color[HTML]{FF0000} {0.7761}}
\\ \hline
\end{tabular}
\vspace{-10pt}
\end{table}
\subsection{Comparison with State-of-the-Art Methods}

\subsubsection{Quantitative Comparison.}
We compare our proposed LCE with other state-of-the-art approaches: DASR \cite{DASR}, IKC \cite{IKC}, DANv1 \cite{DAN}, DANv2 \cite{DANv2}, DCLS \cite{DCLS}, DARSR \cite{DARSR}, KDSR \cite{KDSR}. Blind SR techniques IKC, DANv1, and DANv2 all employ explicit kernel estimation and iterative strategy to improve performance. DCLS also estimates kernels explicitly and employs deep constraint least squares to get clean features. DASR uses contrastive learning to estimate kernels implicitly. DARSR employs the self-supervised method to get a Correction Filter, and KDSR utilizes knowledge distillation for degradation estimation. We utilize pretrained weights whenever available. However, because KDSR is a recent competitive model and no pretrained weights for scale 2 are available, we choose to retrain it using the official implementation.

Tab. \ref{tab 4} shows the quantitative metrics for the isotropic kernel on scales 2 and 4.  Our proposed LCE performs superior to all the other approaches on scales 2 and 4 especially on three large datasets BSD100, Urban100, and Manga109. The iterative procedure in IKC, DANv1, and DANv2 limits their model size and performance. Implicit kernel estimation in DASR also fails to yield a better result. DCLS performs better since it adds deep constraint least squares to get a better result, but its performance is still inferior to our method's. DARSR only utilizes the information in the input LR image and its performance is unstable
under this setting. For scale 4, the input LR images are too small to train DARSR, so we only provide the results in scale 2. The large version KDSR-L uses knowledge distillation to get an implicit degradation estimator. It achieves SOTA
results for scale 4, but our proposed LCE can still get better results for most cases.
 
Tab. \ref{tab 5} presents the quantitative metrics on DIV2KRK and Urban100RK for the anisotropic kernel, which is more complicated to estimate. In this context, our LCE can outperform other approaches for scales 2 and 4. This suggests that LCE can be more beneficial for SR tasks with more complicated anisotropic kernels.
\vspace{-10pt}
\subsubsection{Visual Comparison.} 
\begin{figure*}[!t]
\centering
\includegraphics[width=0.95\textwidth]{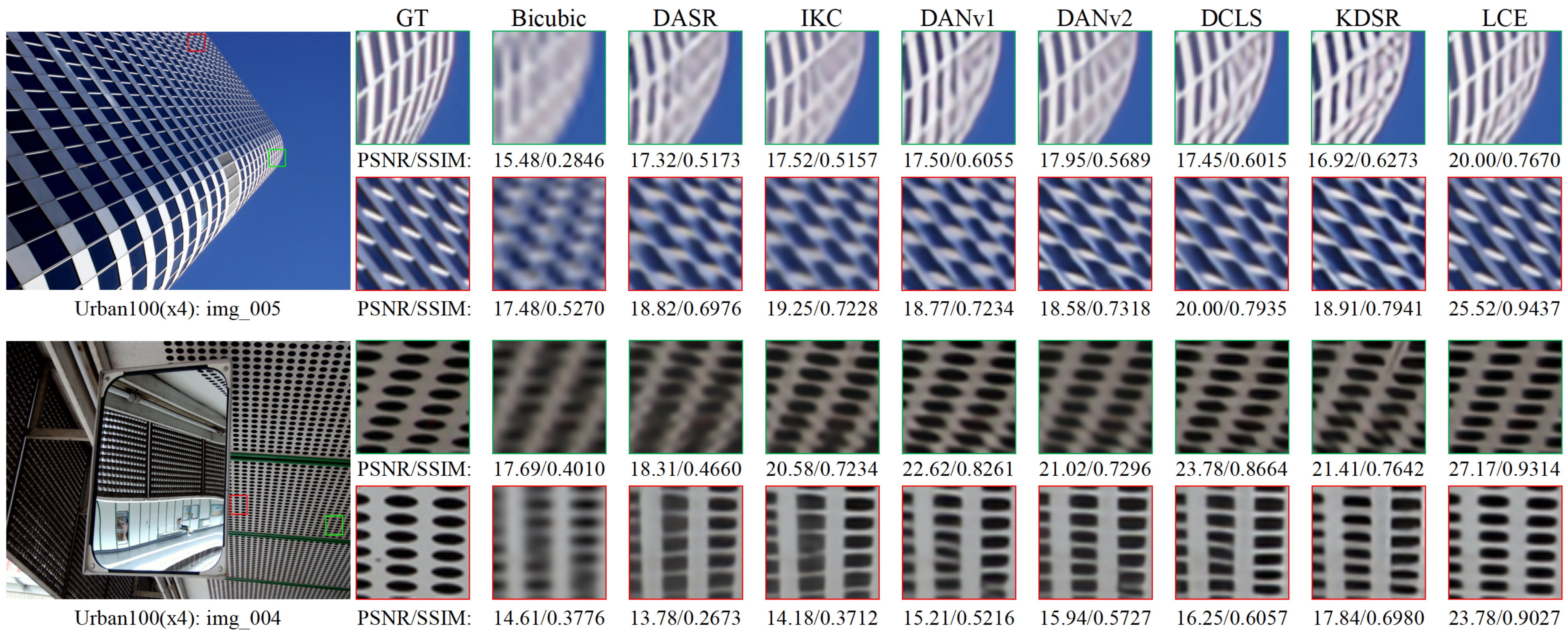}
\caption{Visual results of img\_005 and img\_004 in Urban100 \cite{Urban100}, with scale factor 4 and isotropic kernel width 2.4.}
\label{iso}
\vspace{-10pt}
\end{figure*}

\begin{figure*}[!t]
\centering
\includegraphics[width=0.95\textwidth]{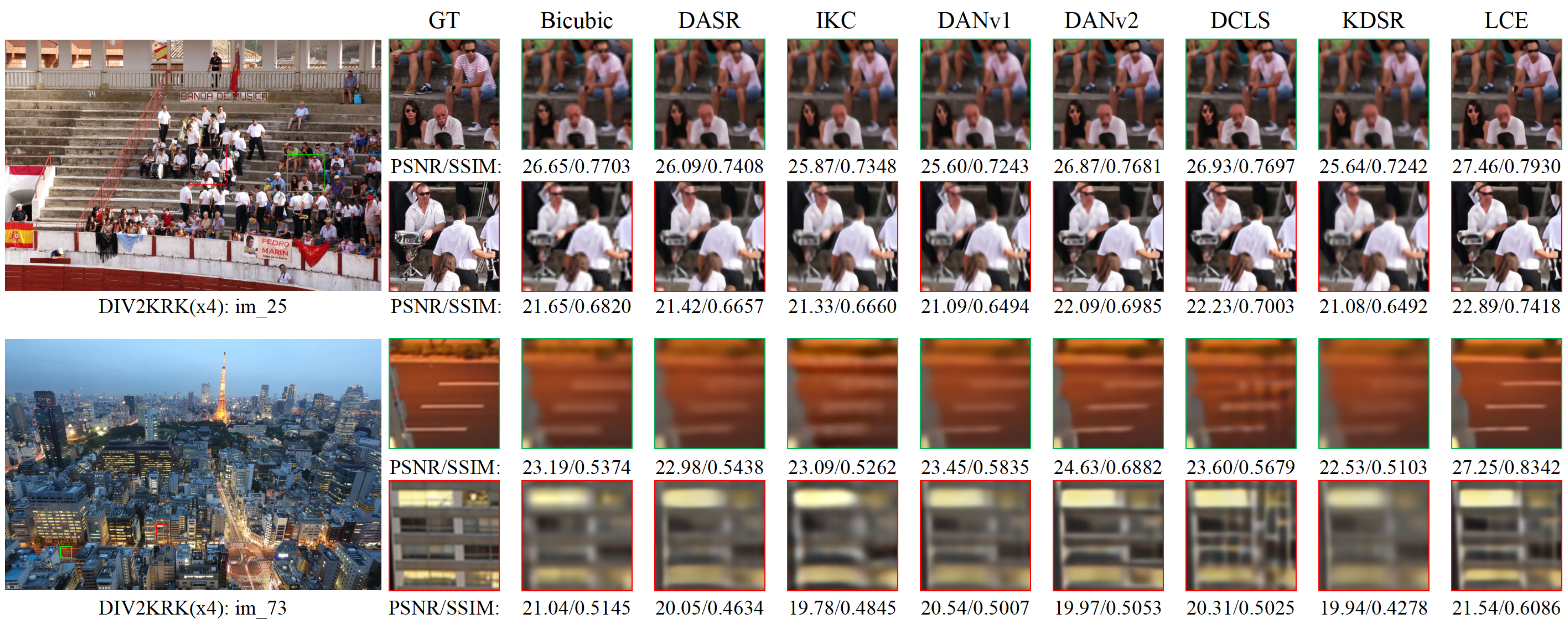}
\caption{Visual results of im\_25 and im\_73 in DIV2KRK \cite{KernelGAN}, with scale factor 4 and anisotropic kernel.}
\label{aniso}
\end{figure*}

The visual results for isotropic kernel and anisotropic kernel are illustrated in Fig. \ref{iso} and Fig. \ref{aniso} respectively. For images in Urban100 and DIV2KRK, we can observe that our methods successfully restore the original textures and edges, while others exhibit severe artifacts or blur. Both Fig. \ref{iso} and Fig. \ref{aniso} demonstrate the effectiveness of our method from the visualization perspective.
\begin{table}[!t]
\setlength{\tabcolsep}{7pt}
\centering
\caption{Complexity analysis about different Blind SR methods. PSNR/SSIM is on Manga109 with isotropic kernel and scale factor 4.}
\label{tab:mult-adds}
\begin{tabular}{cccc}
\hline
Methods & \#Params & Mult-adds(G) & PSNR/SSIM \\
\hline
IKC & 5.32M & 2528.03 & 28.91/0.8782 \\
DANv1 & 4.33M & 1098.33 & 30.50/0.9037 \\
DCLS & 13.63M & 426.43 & 30.86/0.9086 \\
KDSR-L & 14.19M & 623.61 & 30.99/0.9069 \\
LCE(ours) & 14.66M & 894.16 & 31.14/0.9149 \\
\hline
\end{tabular}
\vspace{-10pt}
\end{table}
\vspace{-5pt}
\subsubsection{Complexitiy Analysis.}
We conduct complexity analysis of our LCE and other blind SR methods in Tab. \ref{tab:mult-adds}. The Mult-Adds are calculated on the input size of 180 $\times$320.  We correct the results since Mult-adds and FLOPs are different conceptions. IKC or DANv1 use the iteration strategy so their Mult-adds are much larger compared with their fewer parameters. Our LCE achieves much better results than DCLS and KDSR-L with similar \#Params. Our Mult-adds are also smaller than IKC and DANv1. It indicates our LCE achieves better performance with reasonable computational costs.
\subsection{Experiments on Real Image SR Dataset.}
\begin{figure}[!t]
    \centering
    \includegraphics[width=1\textwidth]{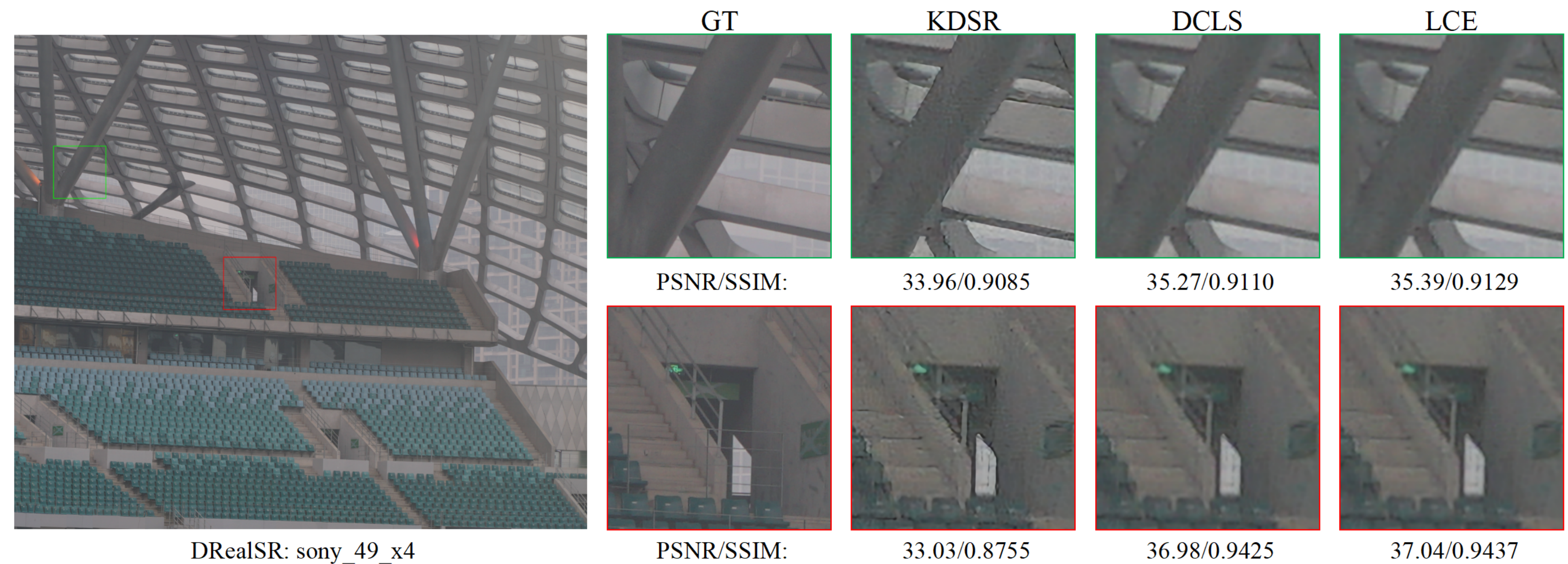}
    \caption{Vision results of real SR on DRealSR \cite{DRealSR} dataset.}
    \label{fig:real_sr}
    \vspace{-5pt}
\end{figure}

We use images in DRealSR\cite{DRealSR} to conduct experiments on real image SR. Images of HR/LR pairs in this dataset are captured with cameras on real scenes and are well-registered using the SIFT method \cite{SIFT}. The degradation process is introduced through actual photography and ISP (Image Signal Processing). We compare with competitive works of KDSR \cite{KDSR} and DCLS \cite{DCLS}. The visual results are shown in Fig. \ref{fig:real_sr}, and the quantitative results are in Tab. \ref{tab:real_sr}. Our proposed LCE achieves better performance on both visual results and quantitative metrics.

\begin{table}[!t]
\centering
\caption{Quantitative comparison of real SR on DRealSR \cite{DRealSR} dataset.}
\setlength{\tabcolsep}{7pt}
\label{tab:real_sr}
\begin{tabular}{ccccc}
\hline
Method & Scale & KDSR \cite{KDSR} & DCLS \cite{DCLS} & LCE(ours) \\
\hline
PSNR & \multirow{2}{*}{×4} & 29.12 & 30.57 & \textbf{30.59} \\

SSIM &  & 0.8216 & 0.8623 & \textbf{0.8676} \\
\hline
\end{tabular}
\vspace{-10pt}
\end{table}
\section{Conclusion}
In this paper, we propose a novel architecture by Learning Correction Errors for Blind SR task. We first obtain a CLR image by using a lightweight Corrector, and then we jointly optimize SR results by utilizing both the original LR image and the frequency learning of the CLR image within an SR network. Additionally, we propose a new Frequency-Self Attention block that combines the mechanisms of frequency spatial attention and self-attention. Extensive experiments demonstrate our method outperforms SOTA blind SR methods in various scenarios.

\bibliographystyle{splncs04}
\bibliography{main}
\end{document}